\documentclass[a4paper, pra, aps, twocolumn, 10pt]{revtex4-2}

\usepackage[utf8x]{inputenc}
\usepackage{amsmath}
\usepackage{amsfonts}
\usepackage{textcomp}
\usepackage{amssymb}
\usepackage{empheq}
\usepackage{esvect}
\usepackage{subcaption}

\usepackage{xcolor}
\definecolor{lightpink}{RGB}{255, 230, 230}
\usepackage{hyperref}
\hypersetup{
  colorlinks   = true, 
  urlcolor     = blue, 
  linkcolor    = blue, 
  citecolor   = red 
}

\newcommand{\tr}{\operatorname{Tr}}
\newcommand{\V}{\mathbb{V}}
\newcommand{\E}{\mathbb{E}}
\newcommand{\CPR}[0]{CPR}
\renewcommand{\O}{\mathcal{O}}
\newcommand{\C}{\mathcal{C}}
\renewcommand{\L}{L_{\rho, \O}}
\newcommand{\p}{\pmb{\phi}}
\newcommand{\wg}{\text{Wg}}


\begin{document}
\author{Nikita~A.~Nemkov}
\affiliation{National University of Science and Technology MISIS, Moscow 119049, Russia}
\email{nnemkov@gmail.com}
    
\title{Statistical models of barren plateaus and anti-concentration of Pauli observables}
\begin{abstract}
We introduce statistical models for each of the three main sources of barren plateaus: non-locality of the observable, entanglement of the initial state, and circuit expressivity. For instance, non-local observables are modeled by random Pauli operators, which lead to barren plateaus with probability exponentially close to one. These models are complementary to the conventional deterministic ones, and often simpler to analyze. Using this framework, we show that in the barren plateau regime any two Pauli observables are anti-concentrated with high probability in the following sense. While each of the observables is localized in an exponentially small parameter subspace, these regions are essentially independent, so that their overlap is yet exponentially smaller than each subspace. This invites to rethink the structure of quantum landscapes with barren plateaus and approaches to their optimization, including warm-start strategies.
\end{abstract}
\maketitle

\section*{Introduction}
Variational quantum algorithms (VQA)~\cite{Cerezo2021} and quantum machine learning (QML) models~\cite{Biamonte2017, Schuld2020} generalize classical optimization and machine learning and have received a great deal of attention in recent years. Initially expected to be useful as simple black-box models for various tasks, they turned out to present many challenges. One of the key issues are trainability problems, mainly associated with poor local minima \cite{Bittel2021, Anschuetz2021, Anschuetz2022} and exponentially flat regions, called barren plateaus (BPs) \cite{McClean2018, Larocca2024}.

In this paper, we focus on the properties of quantum landscapes in the BP regime. When a loss function has a BP, the standard gradient-based and gradient-free optimization methods simply can not escape the flat region and find good solutions \cite{Arrasmith2020}. It is worth noting that VQAs without BPs can typically be efficiently simulated classically \cite{Bermejo2024, Cerezo2023}. Warm-starting optimization in non-BP regions have been proposed as a potential strategy \cite{Skolik2020, Grant2019, Zhang2022, Wang2020, Rudolph2022}. In this work, we will extend the arguments presented in \cite{Nemkov2024} and argue that a typical parameter configuration lying outside the BP region is trivial, because only a single term in the loss function is non-vanishing. This further clarifies the structure of quantum landscapes with BPs and puts additional constraints on what constitutes a useful initialization strategy.

\section*{VQA}
VQAs are defined by initial state $\rho$, observable $\O$, and parameterized quantum circuit (PQC) $U({\pmb{\phi}})$. The corresponding loss function is
\begin{align}
    \L({\pmb{\phi}})=\tr \rho\, U^\dagger({\pmb{\phi}}) \O U({\pmb{\phi}}) \ .
\end{align}
We will sometimes drop $\rho$ or $\O$ as explicit arguments, when the context is clear.

A loss function can be usefully characterized by its mean and variance over the parameter space
\begin{align}
& \E_{\pmb{\phi}}[\L({\pmb{\phi}})] = \int d{\pmb{\phi}}\, \L({\pmb{\phi}})\ ,\\
& \V_{\pmb{\phi}}[\L({\pmb{\phi}})]=\E_{\pmb{\phi}}[\L^2({\pmb{\phi}})]-(\E_{\pmb{\phi}}[\L({\pmb{\phi}})])^2 \ .
\end{align}

VQA is said to have a barren plateau when the variance of its loss function is exponentially small in the number of qubits
\begin{align}
    \V_{\pmb{\phi}}[\L({\pmb{\phi}})] = O(2^{-n}) \ . \label{bp def}
\end{align}
For this definition to make sense one needs to restrict the growth of the observable $\O$ with $n$, and we adopt a convenient condition that $\tr \O^2\le 2^n$, which is satisfied e.g. by a linear combination of Pauli strings $\O=\sum_i c_i P_i$ with $\sum_i |c_i|^2\le1$.

The presence of a BP implies that the loss function can only significantly deviate from its mean value in an exponentially small volume of the parameter space, i.e. it is narrowly concentrated. Three main sources of BPs in noiseless circuits have been identified: circuit expressivity, non-locality of the observable, and high entanglement of the initial state. In addition, quantum noise is known to cause BPs. However, in the case of a sufficiently strong noise the loss is not merely concentrated, but exponentially flat everywhere. Noise-induced BPs are trivial in the context of this work, and will not be considered.

We will focus on a special class of VQAs, which we refer to as Clifford+Pauli rotations (\CPR{}) VQAs. A PQC  $U(\p)$ of a \CPR{} VQA consists only of constant Clifford gates and parameterized Pauli rotations $e^{-\frac{i\phi_k P_k}{2}}$. The initial state $\rho$ is a stabilizer state and the observable $\O$ is a linear combination of $\text{poly}(n)$ Pauli observables. Furthermore, we assume that there are no correlated parameters, i.e. that all $\phi_k\in \p$ are independent.
Except for the latter, these conditions are barely restrictive, since most of the well-studied VQAs satisfy them. The absence of correlated parameters is also common, but does not always hold, e.g. in the case of QAOA \cite{Farhi2014}.

For \CPR{} VQAs the average of any loss function with a single Pauli observable $L_{\rho, P}$ vanishes (or $L_{\rho, P}$ is a constant, and can be disregarded). Moreover, the variance of the loss can be computed as an expectation value over a discrete set of points $\p_c$ \cite{Letcher2023, Nemkov2024}
\begin{multline}
    \V_{\p} [L_{\rho, P}(\p)] = \E_{\p_c} [L_{\rho, P}^2(\p_c)] = \\\frac{1}{|\p_c|}\sum_{\p_c} \left(\tr \rho\, U^\dagger(\p_c) P U(\p_c)\right)^2\ . \label{clifford var}
\end{multline}
Here $\p_c$ is a set of points where every parameter $\phi_k$ is a multiple of $\pi/2$, i.e. $\p_c = \{0, \pi/2, \pi, 3\pi/2\}^{|\p|}$ with $|\p|$ being is the total number of parameters in the circuit. There are $|\p_c|=4^{|\p|}$ such points. We refer to them as Clifford points, because when $\p=\p_c$ the PQC $U(\p_c)$ is a Clifford operator. We emphasize that relation \eqref{clifford var} holds for any \CPR{} VQA and does assume any further properties such as $U(\p)$ being a unitary design.

\section*{Statistical model of barren plateaus}
The sources of BPs listed above are generalizations that may not be entirely accurate. For instance, proving that non-locality induces a BP requires a specific definition of what constitutes a non-local observable and additional restrictions on the PQC, such as being a local unitary 2-design \cite{Cerezo2021a}. Here we propose a different, statistical, model of BPs. We believe it formalizes the same underlying properties, but has different edge-cases to the common BP models. 

Assume that the initial state $\rho$ and the observable $\O$ are drawn from some distributions, and consider the loss function variance averaged over $\rho$ and $\O$. If this expected value is exponentially small
\begin{align}
    \E_\rho \E_\O \left[\V_{\pmb{\phi}}[\L({\pmb{\phi}})]\right] = O(2^{-n}) \ ,
\end{align}
any typical choice of $\rho$ and $\O$ leads to a loss function with a BP with probability exponentially close to one. At the same time, an exponentially small fraction of loss functions from this ensemble may be free of BPs.

\subsection*{Non-locality}
Let us give an explicit example. Consider any \CPR{} VQA with a single-Pauli observable $\O=P$, which is drawn from the ensemble of $n$-qubit random Pauli operators. In this case
\begin{align}
\E_P \left[\V_{\pmb{\phi}}[L_{P}({\pmb{\phi}})]\right] = \E_{\p}\E_P [L^2_{P}(\p)] \ .
\end{align}
Using relation \eqref{clifford var} we can write this as
\begin{multline}
    \E_{\p}\E_P [L^2_{P}(\p)]=\E_{\p_c}\E_P \left(\tr\rho\, U(\p_c)^\dagger P U(\p_c) \right)^2=\\
    \E_P \left(\tr\rho P \right)^2 = 2^{-n} \ .
\end{multline}
Here we discarded $U(\p_c)$ in the second line, because for any fixed Clifford operator $U(\p_c)$ and random Pauli $P$ the combination $U(\p_c)^\dagger P U(\p_c)$ is still a random Pauli, so that $U(\p_c)$ can be discarded when averaging over $P$. The final average is simply the probability that a random Pauli has a non-zero expectation value in a given stabilizer state (e.g. $\rho=|0\rangle^{\otimes n}$), which equals $2^{-n}$.

Hence, a loss function of any \CPR{} VQA with a random Pauli observable has a BP with probability exponentially close to one. At the same time, some loss functions from this ensemble may be free of BPs, e.g. when the PQC is shallow and the Pauli observable happens to be local (which is exponentially unlikely).

We view the ensemble of VQAs with random Pauli observables as a way to formalize the notion of observable non-locality, which is complementary to the existing literature. Our definition does not impose any additional constraints on the PQC, at the cost of only showing that BPs exist with high probability when the observables are sampled from a certain distribution. Importantly, the random Pauli model of non-locality allows succinctly proving the anti-concentration results, which is the main purpose of this work.

\subsection*{Entanglement}
In line with the previous example, we propose to model highly entangled initial states by random stabilizer states. For instance, choosing a reference state $\rho_0 = |0\rangle^{\otimes n}$ a random stabilizer $\rho$ can be sampled by choosing $\rho = \C \rho_0 \C^\dagger$, where $\C$ is a random $n$-qubit Clifford operator. A computation similar to the one we did for random Pauli observables confirms the existence of BPs in this case
\begin{align}
    \E_\rho \E_\phi L^2_\rho(\p)=\E_{\p_c} \E_{\C} \left(\tr \rho_0 \C^\dagger U^\dagger(\p_c) P U(\p_c) \C\right)^2=\\ \E_{\C} \left(\tr \rho_0 \C^\dagger P \C\right)^2 = 2^{-n} \ .
\end{align}
Here we first substituted averaging over $\rho$ with averaging over the Clifford group $\C$, and then used the fact that any $U(\p_c)$ can be discarded in this average.

\subsection*{Expressivity}
The most common measure of circuit expressivity is its proximity to a unitary $t$-design. A PQC $U({\pmb{\phi}})$ is a $t$-design if the ensemble $U({\pmb{\phi}})$ has the same first $t$ moments as the Haar measure $dU$, i.e. the following operator equality holds
\begin{align}
    \int d{\pmb{\phi}} \,\left(U^\dagger({\pmb{\phi}})\right)^{\otimes t} \,  U({\pmb{\phi}})^{\otimes t} =
    \int dU \,\left(U^\dagger\right)^{\otimes t} \,  U^{\otimes t} \ .
\end{align}

A PQC that is a 2-design always leads to a BP \cite{McClean2018} since 
\begin{multline}
\int dU\, \left(\tr \rho U^\dagger \O U\right)^2=\\\tr \O^2 \left(\frac{(\tr \rho)^2}{4^n-1}-\frac{\tr\rho^2}{2^n(4^n-1)}\right)=O(2^{-n}) \ .
\end{multline}

When the circuit is an approximate instead of the exact 2-design, as is usually the case in practice, this relation also holds approximately \cite{Holmes2021}. For clarity, in the following we always assume that unitary designs are exact.

While expressivity is meant to capture the intuition that a given PQC covers a large portion of the Hilbert space, the 2-design property in addition implies that this covering is uniform and unbiased to some degree. It may be useful to lift this restriction by making the expressivity model statistical. One possibility is to consider such PQC $U({\pmb{\phi}})$, that when evaluated at random Clifford points ${\pmb{\phi}}_c$ sample uniformly from the Clifford group $U({\pmb{\phi}}_c)\sim\C$. The presence of BPs in this case is confirmed by a simple computation
\begin{multline}
\E_{\pmb{\phi}} \left[\L^2({\pmb{\phi}})\right]=\E_{{\pmb{\phi}}_c} \left[\left(\tr \rho_0 U({\pmb{\phi}}_c) ^\dagger \O U({\pmb{\phi}}_c)\right)^2\right]= \\ 
\E_{\C} \left[\left(\tr \rho_0 \C \O \C)\right)^2\right]=O(2^{-n}) \ .
\end{multline}

This definition, however, still requires uniformity of sampling from the Clifford group. A further possibility is to consider an ensemble of unitary circuits $\mathcal{U}$ such that the uniform sampling over the Clifford group is achieved only when both $U$ and $\p_c$ at drawn at random. At the moment, we do not have a natural example of such a distribution over circuits. Our main anti-concentration result will be proven in the conventional model of expressivity-induced BPs.

\section*{Anti-concentration of Pauli observables}
Generally, an observable in VQA can be decomposed into a linear combination of Pauli operators $\O=\sum_i c_i P_i$, which implies that the full loss function is also a linear combination of single-Pauli loss functions $L_{\O}=\sum_i c_i L_{P_i}$. 

For any non-constant single-Pauli loss function $L_{P}$ its max and min values are $L_P=\pm 1$, attained, e.g., at a subset of Clifford points. If, in addition, $L_{P}$ has a BP, it must be highly concentrated in the regions containing these Clifford points. An important question is whether the loss functions corresponding to different Pauli observables tend to concentrate near the same or different Clifford points. We will refer to the latter case as anti-concentration of Pauli observables.

Anti-concentration of Pauli observables is likely to be an undesirable effect from the perspective of VQAs. As noted in \cite{Nemkov2024}, it creates many poor local minima in the landscape of the full loss function $L_{\O}$. Moreover, it shows that even within the exponentially small volume where the full loss function is concentrated, most points are trivial, because they correspond to a single non-zero Pauli term. Hence, the subspace of interesting solutions is yet exponentially smaller than the localization region of the full loss function.

To quantify whether the non-zero Clifford points of two Pauli observables $P_1, P_2$ are correlated, consider the following expectation value
\begin{align}
A_{P_1, P_2}=\E_{{\pmb{\phi}}_c} [|L_{P_1}({\pmb{\phi}}_c)L_{P_2}({\pmb{\phi}}_c)|]=\E_{{\pmb{\phi}}_c} [L^2_{P_1}({\pmb{\phi}}_c)L^2_{P_2}({\pmb{\phi}}_c)] \ . \label{c def}
\end{align}
This correlator counts the fraction of Clifford points where both $L_{P_1}$ and $L_{P_2}$ are non-zero. When the two functions are fully correlated, i.e. non-zero at the same points, we have $A_{P_1, P_2}=O(2^{-n})$. In the opposite case, when overlap is fully random and Pauli observables anti-concentrate, we expect $A_{P_1, P_2}=O(4^{-n})$. 

Note that the absolute value $|L_{P}|$ (or a square $|L_{P}|^2$) is necessary in definition \eqref{c def}, because the signed expectation value can turn out zero even when the overlap between non-zero points of $P_1$ and $P_2$ is large.

\subsection*{Non-locality}
Anti-concentration is simplest to prove in our model of non-locality. Consider the average of $A_{P_1, P_2}$ in the ensemble where $P_1$ and $P_2$ are independent random Pauli operators
\begin{multline}
    \E_{P_1} \E_{P_2}[A_{P_1, P_2}]= \E_{{\pmb{\phi}}_c}\big[\E_{P_1}[L^2_{P_1}({\pmb{\phi}}_c)]\,\, \E_{P_2}[L^2_{P_2}({\pmb{\phi}}_c)]\big]=\\\E_{{\pmb{\phi}}_c}[2^{-n}\times 2^{-n}]=4^{-n} \ .
\end{multline}
We used the independence of $P_1$ and $P_2$ to factor the expectation value. This simple calculation shows that any typical pair of random observables $P_1, P_2$ is anti-concentrated with probability exponentially close to one. 

\subsection*{Entanglement}
Now we consider the average of correlator $A_{P_1, P_2}$ over random stabilizer states for two arbitrary Pauli operators
\begin{multline}
    \E_\rho [A_{P_1, P_2}] = \\ \E_{{\pmb{\phi}}_c} \E_\C [(\tr \rho_0 \C^\dagger U^\dagger_c P_1 U_c \C)^2 (\tr \rho_0 \C^\dagger U^\dagger_c P_2 U_c \C)^2] = \\
    \E_\C [(\tr \rho_0 \C^\dagger P_1 \C)^2 (\tr \rho_0 \C^\dagger P_2\C)^2] \ . \label{A ent}
\end{multline}
Here we first expressed averaging over the stabilizer states as averaging over the Clifford gates $\C$, and then used the fact that $U_c=U({\pmb{\phi}}_c)$ are also Clifford gates, and hence do not affect the average.

Consider a subset of the Clifford gates $\C_1$ that have the property that $(\tr \rho_0 \C^\dagger P_1 \C)^2=1$. For $\rho_0=|0\rangle^{\otimes n}$ these are the gates that transform $P_1$ into a Pauli operator of the form $C_1^\dagger P_1 C_1=Z^{\alpha}=Z_1^{\alpha_1}\dots Z_n^{\alpha_n}$, i.e. free of $X$ factors. The average then reduces to
\begin{align}
    \E_\rho [A_{P_1, P_2}] = \frac{|\C_1|}{|\C|} \E_{\C_1} (\tr \rho_0 \C^\dagger P_2\C)^2 \ . \label{avg C1}
\end{align}
The ratio of dimensions
\begin{align}
 \frac{|\C_1|}{|\C|}=\frac{2^n-1}{4^n-1}  
\end{align}
can be computed as the number of all Pauli gates free of $X$ factors and reachable from $P_1$, which is $2^n-1$ (excluding the identity), divided by the number of all Paulis reachable from $P_1$, which is $4^n-1$.

Now introduce $\C_{1, \alpha}$, a subset of $\C_1$ that transforms $P_1$ into $Z^{\alpha}$ with some particular $\alpha$. From the properties of the Clifford group it can be shown, that the orbit $\mathcal{P}_{\alpha}=\C_{1,\alpha}^\dagger P_2 \C_{1,\alpha}$ uniformly represents all Pauli operators, which commute with $Z^{\alpha}$ if $[P_1,P_2]=0$, and anti-commute with $Z^{\alpha}$ otherwise. Hence, if $P_1$ anti-commutes with $P_2$ the average \eqref{avg C1} is exactly zero, because any Pauli anti-commuting with $Z^{\alpha}$ must have a non-trivial $X$ factor and thus vanishing expectation value in $\rho_0$.

Finally, consider average $\E_{\C_{1,\alpha}}[(\tr \rho_0 \C^\dagger P_2\C)^2]$. It is equal to the fraction of Pauli operators from the orbit $\mathcal{P}_{\alpha}$ having non-zero expectation value in $\rho_0$. For any $\alpha$, there are exactly $|\gamma|=2^{n-1}$ Pauli operators of the form $X^{\gamma}$ such that $[X^{\gamma}, Z^{\alpha}]=0$. Therefore, all elements of $\mathcal{P}_{\alpha}$ must be of the form $Z^{\beta} X^{\gamma}$. There are in total $|\beta||\gamma|-2=2^{2n-1}-2$ such operators (excluding the identity and $Z^\alpha$ itself, which are not reachable from $P_2\neq P_1$). Among them, only $|\beta|-2$ operators of the form $Z^{\beta}$ have non-zero expectation value in $\rho_0$ (again excluding $Z^{\alpha}$ and the identity). Therefore 
\begin{align}
\E_{C_{1,\alpha}} (\tr \rho_0 \C^\dagger P_2\C)^2=\frac{2^n-2}{2^{2n-1}-2} \ .
\end{align}
Since all subsets $\C_{1, \alpha}$ have equal size we find
\begin{align}
\E_\rho [A_{P_1, P_2}]=\frac{2^n-1}{4^n-1}\frac{2^n-2}{2^{2n-1}-2}=O(4^{-n})
\end{align}
for any two operators $P_1, P_2$ that are not equal to each other or the identity. This proves the anti-concentration in our model of the entanglement-induced BPs.

\subsection*{Expressivity}
If we assume a statistical model of expressivity-induced BPs introduced above, where simultaneously sampling from the ensemble of unitaries and Clifford angles $U(\p_c)$ is equivalent to sampling from the Clifford group, the average of expectation value \eqref{c def} can be reduced to the one we computed for the entanglement-induced BPs
\begin{multline}
    \E_U [A_{P_1, P_2}]=\\\E_\C [(\tr \rho_0 \C^\dagger P_1 \C)^2 (\tr \rho_0 \C^\dagger P_2\C)^2] = O(4^{-n})\ .
\end{multline}

In this case, however, we can also consider a continuous counterpart of \eqref{c def}
\begin{multline}
    A'_{P_1, P_2}=\E_{\pmb{\phi}} \left[L^2_{P_1}(\p)L^2_{P_2}(\p)\right] =\\
    \E_{\p} \left[\left(\tr \rho\, U^\dagger(\p) P_1 U(\p)\right)^2 \left(\tr \rho\, U^\dagger(\p) P_2 U(\p)\right)^2\right]\ . \label{c' def}
\end{multline}
If $U(\p)$ is a 4-design, this expectation value can be evaluated explicitly using Weingarten calculus (see e.g. \cite{Puchaa2011})
\begin{multline}
    \int dU\,\,U_{i_1, j_1}\dots U_{i_t, j_t}\,\,\overline{U}_{i_1', j_1'}\dots \overline{U}_{i_t', j_t'} = \\ \sum_{\sigma, \tau} \delta_{i_1, i_{\sigma(1)}}\cdots \delta_{i_n, i_{\sigma(n)}}\,\, \delta_{j_1, j_{\tau(1)}}\cdots \delta_{j_n, j_{\tau(n)}} \wg (\tau \sigma^{-1}, 2^{n}) \ . \label{wg}
\end{multline}
Here $U_{i, j}$ are matrix elements of $U$, $\overline{U}_{i, j}$ their complex conjugates, $\sigma$ and $\tau$ run over all permutations of $t$ elements, and $\wg$ is the Weingarten function, which introduces a weight for each term in this sum.

Applying \eqref{wg} to \eqref{c' def} generates many terms which are products of fully contracted tensors $\rho^{\otimes 4}$ and $P_1^{\otimes 2}P_2^{\otimes 2}$. The majority of these terms are zero, however, because most contractions of the Pauli terms are zero, e.g. $\tr P_1=\tr P_1^2 P_2=\tr P_1 P_2=0$. The only two non-vanishing contractions are $\tr P_1^2 P_2^2=2^n$, corresponding to $\tau = (1234)$, and $\tr P_1^2 \tr P_2^2=4^n$, corresponding to $\tau=(12)(34)$. Each of these terms multiplies a non-trivial amount of contractions of tensor $\rho^{\otimes 4}$, but for our purposes it is not necessary to enlist all of them. The Weingarten function of order 4, which asymptotically dominates others as $n\to\infty$, corresponds to the identity permutation and equals
\begin{multline}
    \wg(\text{id}, 2^n)=\frac{N^4-8N^2+6}{N^2(N^2-1)(N^2-4)(N^2-9)}=\\ 16^{-n}\left(1+O(2^{-n})\right)\ ,
\end{multline}
where $N=2^n$. Therefore, the leading term in the asymptotic expansion of \eqref{c' def} is
\begin{multline}
    A'_{P_1, P_2}=\\\tr P_1^2 \tr P_2^2 (\tr \rho^2)^2 \wg(\text{id}, 2^n)\left(1+O(2^{-n})\right)\\ = 4^{-n}\left(1+O(2^{-n})\right) \ ,
\end{multline}
which establishes a continuous version of Pauli anti-concentration for 4-designs.

While our proof does not apply to 2-designs, the anti-concentration result may still be valid. Note that we showed that a similar average over the Clifford group \eqref{A ent} anti-concentrates, despite the Clifford group not being a 4-design \cite{Webb2015, Zhu2015, Zhu2016}. If, however, the gap is real, there is an interesting possibility that circuits that are 2-designs but not 4-designs may strike a delicate balance between being classically simulable and being too expressive to cause Pauli anti-concentration. 

\begin{figure}

\begin{subfigure}[b]{\linewidth}
   \includegraphics[width=0.8\linewidth]{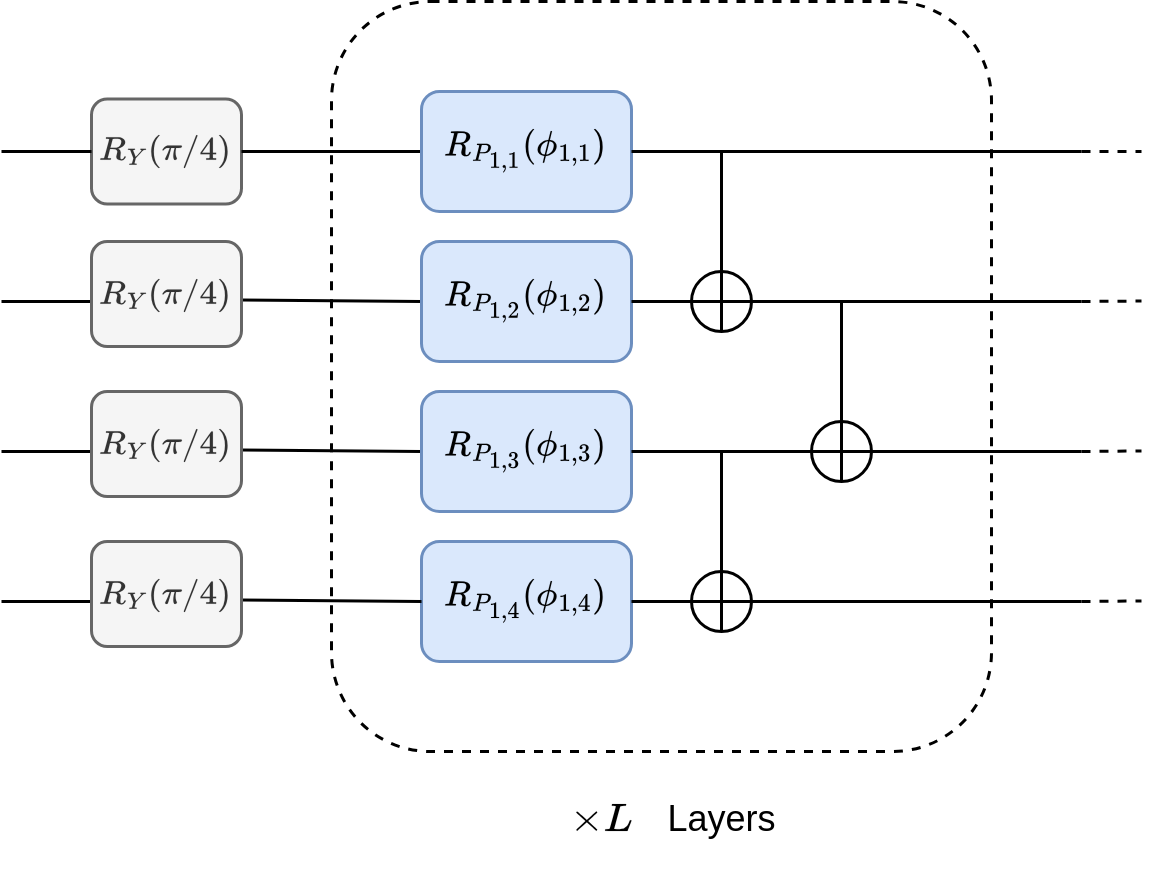}
   \caption{Example of a HEA circuit with $n=4$ qubits. The axes in Pauli rotation operators $R_{P_{i, j}}$ are chosen randomly.}
   \label{fig hea} 
\end{subfigure}

\begin{subfigure}[b]{\linewidth}
   \includegraphics[width=\linewidth]{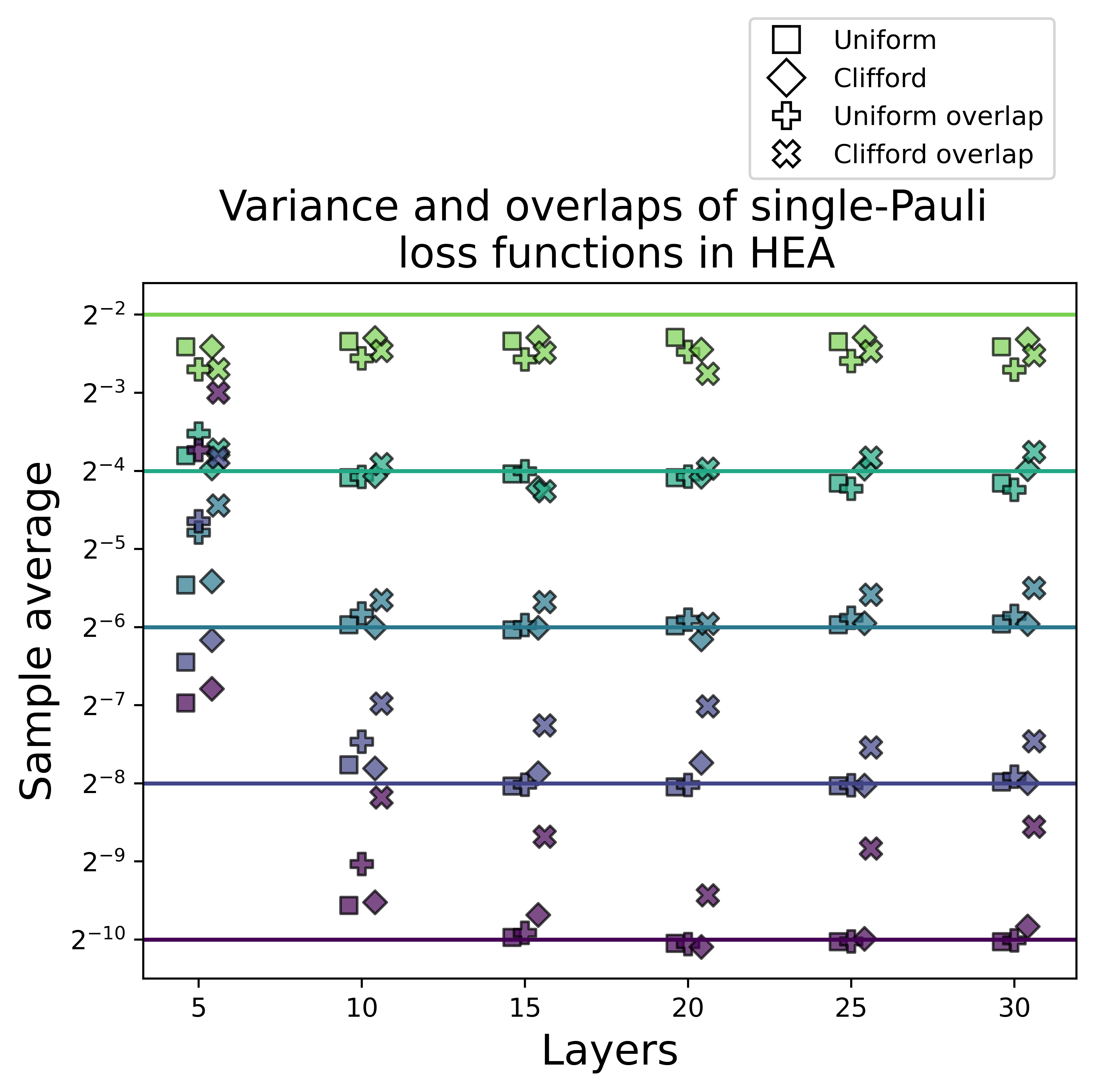}
   \caption{The plot shows: (i) loss function variances computed by uniform $\square=\E_{\p}[L^2_P(\phi)]$ and Clifford \includegraphics[height=\fontcharht\font`\B]{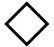}$=\E_{\p_c}[L^2_P(\phi_c)]$ sampling (ii) estimates of continuous \includegraphics[height=\fontcharht\font`\B]{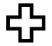}$=A_{P_1, P_2}$ and discrete \includegraphics[height=\fontcharht\font`\B]{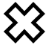}$= A_{P_1, P_2}$ correlators. Loss function values $L_{P_i}$ are computed at 500 $\p$ samples (drawn from the uniform distribution or Clifford points $\p_c$ in the discrete case) for all two-body nearest-neighbor Pauli terms existing for a given number of qubits. This data is then used to compute the loss function variances of and overlap correlators. The figure shows the variances averaged over all Pauli observables and correlators averaged over all pairs of Pauli observables. The correlators $A_{P_1, P_2}$ and $A'_{P_1, P_2}$ are multiplied by $2^n$ before plotting, to match the expected scale of the loss variances.}
   \label{fig plot}
\end{subfigure}

\end{figure}

We probe this question numerically using hardware-efficient ansatz (HEA) \cite{Kandala2017} (see Fig.~\ref{fig hea}), which was shown in \cite{Liu2022c} to interpolate between a 2-design and higher-order designs as the number of layers $L$ increases. For circuits with up to $n=10$ qubits and $L=30$ layers we estimate loss function variances by sampling from 500 random points for both discrete $A_{P_1, P_2}$ and continuos $A'_{P_1, P_2}$ anti-concentration correlators for all two-body nearest-neighbor Pauli operators, of which there are $9(n-1)$. Results are reported in Fig.~\ref{fig plot} and are somewhat inconclusive. We see that the loss function variance approaches the expected value of $2^{-n}$ both when computed via the uniform and Clifford sampling. The discrete $A_{P_1, P_2}$ and continuous $A'_{P_1, P_2}$ correlators also converge to the expected values $4^{-n}$ at a similar rate as the loss variance, indicating that in this case the onset of anti-concentration happens simultaneously with the onset of BPs. The convergence of the discrete correlators $A_{P_1, P_2}$ at $n=8$ and $n=10$ appears to be slightly slower, but it is not clear if the effect is significant. These numerical experiments quickly get very challenging as $n$ grows. This is not only because the complexity of quantum circuit simulation increases, but also because the estimated values decay exponentially. Hence, to keep a reasonable estimation precision, the number of samples needs to exponentially increase as well. Extending these experiments to large $n$ would require advanced simulation techniques and goes beyond the scope of this work. Our code relies on PennyLane \cite{Bergholm2018} and is available at \cite{bp_hea}.

\section*{Conclusion}
In this work, we introduced several statistical models of barren plateaus, that are complementary to the conventional ones. Using these models, we showed that in all three scenarios where BPs arise, i.e., non-locality of the observable, entanglement of the initial state, and circuit expressivity, Pauli observables anti-concentrate with probability exponentially close to one. This implies that the typical landscapes with BPs may be even more challenging than previous believed. Not only do individual Pauli observables suffer from concentration in an exponentially small volume of the parameter space, but their respective regions of concentration are likely to be essentially independent of each other. This implies the non-flat regions of the full loss functions are dominated by parameter patches where only a single Pauli term is not exponentially small. In other words, parameter configurations where any non-trivial number of Pauli terms are simultaneously off their respective BPs constitute only an exponentially small portion of an already exponentially small localization subspace. While this does not imply that non-trivial optimization solutions do not exist at all, it further highlights that the standard methods to look for such solutions, e.g. those based on gradient descent, are inadequate. Interestingly, one may try to look for non-trivial solutions in the vicinity of those Clifford points, where an exceptionally large number of Pauli operators are simultaneously non-zero. This would turn the warm-start into a discrete optimization problem, which is interesting to consider. We believe that further investigation of the quantum loss landscapes along these lines may be very fruitful.

\subsection*{Acknowledgments}
N.A.N. thanks the Russian Science Foundation Grant No. 23-71-01095 (theoretical results). 
Numerical experiments are supported by the Priority 2030 program at the National University of Science and Technology ``MISIS'' under the project K1-2022-027.

\bibliographystyle{quantum}
\bibliography{refs.bib}

\begin{thebibliography}{10}

\bibitem{Cerezo2021}
M.~Cerezo, Andrew Arrasmith, Ryan Babbush, Simon~C. Benjamin, Suguru Endo, Keisuke Fujii, Jarrod~R. McClean, Kosuke Mitarai, Xiao Yuan, Lukasz Cincio, and Patrick~J. Coles.
\newblock ``{Variational quantum algorithms}''.
\newblock \href{https://dx.doi.org/10.1038/s42254-021-00348-9}{Nature Reviews Physics 2021 3:9 {\bf 3}, 625--644}~(2021).
\newblock  \href{http://arxiv.org/abs/2012.09265}{arXiv:2012.09265}.

\bibitem{Biamonte2017}
Jacob Biamonte, Peter Wittek, Nicola Pancotti, Patrick Rebentrost, Nathan Wiebe, and Seth Lloyd.
\newblock ``{Quantum machine learning}''.
\newblock \href{https://dx.doi.org/10.1038/nature23474}{Nature {\bf 549}, 195--202}~(2017).
\newblock  \href{http://arxiv.org/abs/1611.09347}{arXiv:1611.09347}.

\bibitem{Schuld2020}
Maria Schuld, Ryan Sweke, and Johannes~Jakob Meyer.
\newblock ``{The effect of data encoding on the expressive power of variational quantum machine learning models}''.
\newblock \href{https://dx.doi.org/10.1103/PhysRevA.103.032430}{Physical Review A{\bf 103}}~(2020).
\newblock  \href{http://arxiv.org/abs/2008.08605v2}{arXiv:2008.08605v2}.

\bibitem{Bittel2021}
Lennart Bittel and Martin Kliesch.
\newblock ``{Training Variational Quantum Algorithms Is NP-Hard}''.
\newblock \href{https://dx.doi.org/10.1103/PhysRevLett.127.120502}{Physical Review Letters {\bf 127}, 120502}~(2021).
\newblock  \href{http://arxiv.org/abs/2101.07267}{arXiv:2101.07267}.

\bibitem{Anschuetz2021}
Eric~R. Anschuetz.
\newblock ``{Critical Points in Quantum Generative Models}''~(2021).
\newblock  \href{http://arxiv.org/abs/2109.06957}{arXiv:2109.06957}.

\bibitem{Anschuetz2022}
Eric~R. Anschuetz and Bobak~T. Kiani.
\newblock ``{Quantum variational algorithms are swamped with traps}''.
\newblock \href{https://dx.doi.org/10.1038/s41467-022-35364-5}{Nature Communications {\bf 13}, 7760}~(2022).
\newblock  \href{http://arxiv.org/abs/2205.05786}{arXiv:2205.05786}.

\bibitem{McClean2018}
Jarrod~R. McClean, Sergio Boixo, Vadim~N. Smelyanskiy, Ryan Babbush, and Hartmut Neven.
\newblock ``{Barren plateaus in quantum neural network training landscapes}''.
\newblock \href{https://dx.doi.org/10.1038/s41467-018-07090-4}{Nature Communications {\bf 9}, 1--7}~(2018).
\newblock  \href{http://arxiv.org/abs/1803.11173}{arXiv:1803.11173}.

\bibitem{Larocca2024}
Mart{\'{i}}n Larocca, Supanut Thanasilp, Samson Wang, Kunal Sharma, Jacob Biamonte, Patrick~J. Coles, Lukasz Cincio, Jarrod~R. McClean, Zo{\"{e}} Holmes, and M.~Cerezo.
\newblock ``{Barren plateaus in variational quantum computing}''.
\newblock \href{https://dx.doi.org/10.1038/s42254-025-00813-9}{Nature Reviews Physics {\bf 7}, 174--189}~(2025).
\newblock  \href{http://arxiv.org/abs/2405.00781}{arXiv:2405.00781}.

\bibitem{Arrasmith2020}
Andrew Arrasmith, M.~Cerezo, Piotr Czarnik, Lukasz Cincio, and Patrick~J. Coles.
\newblock ``{Effect of barren plateaus on gradient-free optimization}''.
\newblock \href{https://dx.doi.org/10.22331/q-2021-10-05-558}{Quantum {\bf 5}, 558}~(2021).
\newblock  \href{http://arxiv.org/abs/2011.12245v2}{arXiv:2011.12245v2}.

\bibitem{Bermejo2024}
Pablo Bermejo, Paolo Braccia, Manuel~S. Rudolph, Zo{\"{e}} Holmes, Lukasz Cincio, and M.~Cerezo.
\newblock ``{Quantum Convolutional Neural Networks are (Effectively) Classically Simulable}''~(2024).
\newblock  \href{http://arxiv.org/abs/2408.12739}{arXiv:2408.12739}.

\bibitem{Cerezo2023}
M.~Cerezo, Martin Larocca, Diego Garc{\'{i}}a-Mart{\'{i}}n, N.~L. Diaz, Paolo Braccia, Enrico Fontana, Manuel~S. Rudolph, Pablo Bermejo, Aroosa Ijaz, Supanut Thanasilp, Eric~R. Anschuetz, and Zo{\"{e}} Holmes.
\newblock ``{Does provable absence of barren plateaus imply classical simulability? Or, why we need to rethink variational quantum computing}''~(2023).
\newblock  \href{http://arxiv.org/abs/2312.09121}{arXiv:2312.09121}.

\bibitem{Skolik2020}
Andrea Skolik, Jarrod~R. McClean, Masoud Mohseni, Patrick van~der Smagt, and Martin Leib.
\newblock ``{Layerwise learning for quantum neural networks}''.
\newblock \href{https://dx.doi.org/10.1007/s42484-020-00036-4}{Quantum Machine Intelligence{\bf 3}}~(2020).
\newblock  \href{http://arxiv.org/abs/2006.14904v1}{arXiv:2006.14904v1}.

\bibitem{Grant2019}
Edward Grant, Leonard Wossnig, Mateusz Ostaszewski, and Marcello Benedetti.
\newblock ``{An initialization strategy for addressing barren plateaus in parametrized quantum circuits}''.
\newblock \href{https://dx.doi.org/10.22331/q-2019-12-09-214}{Quantum {\bf 3}, 214}~(2019).
\newblock  \href{http://arxiv.org/abs/1903.05076v3}{arXiv:1903.05076v3}.

\bibitem{Zhang2022}
Xiao-Ming Zhang, Tongyang Li, and Xiao Yuan.
\newblock ``{Quantum State Preparation with Optimal Circuit Depth: Implementations and Applications}''~(2022).
\newblock  \href{http://arxiv.org/abs/2201.11495}{arXiv:2201.11495}.

\bibitem{Wang2020}
Samson Wang, Enrico Fontana, M.~Cerezo, Kunal Sharma, Akira Sone, Lukasz Cincio, and Patrick~J. Coles.
\newblock ``{Noise-induced barren plateaus in variational quantum algorithms}''.
\newblock \href{https://dx.doi.org/10.1038/s41467-021-27045-6}{Nature Communications {\bf 12}, 6961}~(2021).
\newblock  \href{http://arxiv.org/abs/2007.14384}{arXiv:2007.14384}.

\bibitem{Rudolph2022}
Manuel~S. Rudolph, Jacob Miller, Danial Motlagh, Jing Chen, Atithi Acharya, and Alejandro Perdomo-Ortiz.
\newblock ``{Synergy Between Quantum Circuits and Tensor Networks: Short-cutting the Race to Practical Quantum Advantage}''~(2022).
\newblock  \href{http://arxiv.org/abs/2208.13673}{arXiv:2208.13673}.

\bibitem{Nemkov2024}
Nikita~A. Nemkov, Evgeniy~O. Kiktenko, and Aleksey~K. Fedorov.
\newblock ``{Barren plateaus swamped with traps}''.
\newblock \href{https://dx.doi.org/10.1103/PhysRevA.111.012441}{Physical Review A {\bf 111}, 012441}~(2025).
\newblock  \href{http://arxiv.org/abs/2405.05332}{arXiv:2405.05332}.

\bibitem{Farhi2014}
Edward Farhi, Jeffrey Goldstone, and Sam Gutmann.
\newblock ``{A Quantum Approximate Optimization Algorithm}''~(2014).
\newblock  \href{http://arxiv.org/abs/1411.4028}{arXiv:1411.4028}.

\bibitem{Letcher2023}
Alistair Letcher, Stefan Woerner, and Christa Zoufal.
\newblock ``{Tight and Efficient Gradient Bounds for Parameterized Quantum Circuits}''~(2023).
\newblock  \href{http://arxiv.org/abs/2309.12681}{arXiv:2309.12681}.

\bibitem{Cerezo2021a}
M.~Cerezo, Akira Sone, Tyler Volkoff, Lukasz Cincio, and Patrick~J. Coles.
\newblock ``{Cost function dependent barren plateaus in shallow parametrized quantum circuits}''.
\newblock \href{https://dx.doi.org/10.1038/s41467-021-21728-w}{Nature Communications 2021 12:1 {\bf 12}, 1--12}~(2021).
\newblock  \href{http://arxiv.org/abs/2001.00550}{arXiv:2001.00550}.

\bibitem{Holmes2021}
Zo{\"{e}} Holmes, Kunal Sharma, M.~Cerezo, and Patrick~J. Coles.
\newblock ``{Connecting Ansatz Expressibility to Gradient Magnitudes and Barren Plateaus}''.
\newblock \href{https://dx.doi.org/10.1103/PRXQuantum.3.010313}{PRX Quantum {\bf 3}, 010313}~(2022).
\newblock  \href{http://arxiv.org/abs/2101.02138v2}{arXiv:2101.02138v2}.

\bibitem{Puchaa2011}
Zbigniew Pucha{\l}a and J.A. Miszczak.
\newblock ``{Symbolic integration with respect to the Haar measure on the unitary groups}''.
\newblock \href{https://dx.doi.org/10.1515/bpasts-2017-0003}{Bulletin of the Polish Academy of Sciences Technical Sciences {\bf 65}, 21--27}~(2017).
\newblock  \href{http://arxiv.org/abs/1109.4244v2}{arXiv:1109.4244v2}.

\bibitem{Webb2015}
Zak Webb.
\newblock ``{The Clifford group forms a unitary 3-design}''.
\newblock \href{https://dx.doi.org/10.26421/QIC16.15-16-8}{Quantum Information and Computation {\bf 16}, 1379--1400}~(2016).
\newblock  \href{http://arxiv.org/abs/1510.02769}{arXiv:1510.02769}.

\bibitem{Zhu2015}
Huangjun Zhu.
\newblock ``{Multiqubit Clifford groups are unitary 3-designs}''.
\newblock \href{https://dx.doi.org/10.1103/PhysRevA.96.062336}{Physical Review A{\bf 96}}~(2015).
\newblock  \href{http://arxiv.org/abs/1510.02619}{arXiv:1510.02619}.

\bibitem{Zhu2016}
Huangjun Zhu, Richard Kueng, Markus Grassl, and David Gross.
\newblock ``{The Clifford group fails gracefully to be a unitary 4-design}''~(2016).
\newblock  \href{http://arxiv.org/abs/1609.08172}{arXiv:1609.08172}.

\bibitem{Kandala2017}
Abhinav Kandala, Antonio Mezzacapo, Kristan Temme, Maika Takita, Markus Brink, Jerry~M. Chow, and Jay~M. Gambetta.
\newblock ``{Hardware-efficient Variational Quantum Eigensolver for Small Molecules and Quantum Magnets}''.
\newblock \href{https://dx.doi.org/10.1038/nature23879}{Nature {\bf 549}, 242--246}~(2017).
\newblock  \href{http://arxiv.org/abs/1704.05018}{arXiv:1704.05018}.

\bibitem{Liu2022c}
Minzhao Liu, Junyu Liu, Yuri Alexeev, and Liang Jiang.
\newblock ``{Estimating the randomness of quantum circuit ensembles up to 50 qubits}''.
\newblock \href{https://dx.doi.org/10.1038/s41534-022-00648-7}{npj Quantum Information {\bf 8}, 137}~(2022).
\newblock  \href{http://arxiv.org/abs/2205.09900}{arXiv:2205.09900}.

\bibitem{Bergholm2018}
Ville Bergholm, Josh Izaac, Maria Schuld, Christian Gogolin, Shahnawaz Ahmed, Vishnu Ajith, M.~Sohaib Alam, Guillermo Alonso-Linaje, B.~AkashNarayanan, Ali Asadi, Juan~Miguel Arrazola, Utkarsh Azad, Sam Banning, Carsten Blank, Thomas~R Bromley, Benjamin~A. Cordier, Jack Ceroni, Alain Delgado, Olivia {Di Matteo}, Amintor Dusko, Tanya Garg, Diego Guala, Anthony Hayes, Ryan Hill, Aroosa Ijaz, Theodor Isacsson, David Ittah, Soran Jahangiri, Prateek Jain, Edward Jiang, Ankit Khandelwal, Korbinian Kottmann, Robert~A. Lang, Christina Lee, Thomas Loke, Angus Lowe, Keri McKiernan, Johannes~Jakob Meyer, J.~A. Monta{\~{n}}ez-Barrera, Romain Moyard, Zeyue Niu, Lee~James O'Riordan, Steven Oud, Ashish Panigrahi, Chae-Yeun Park, Daniel Polatajko, Nicol{\'{a}}s Quesada, Chase Roberts, Nahum S{\'{a}}, Isidor Schoch, Borun Shi, Shuli Shu, Sukin Sim, Arshpreet Singh, Ingrid Strandberg, Jay Soni, Antal Sz{\'{a}}va, Slimane Thabet, Rodrigo~A. Vargas-Hern{\'{a}}ndez, Trevor Vincent, Nicola Vitucci, Maurice Weber, David Wierichs,
  Roeland Wiersema, Moritz Willmann, Vincent Wong, Shaoming Zhang, and Nathan Killoran.
\newblock ``{PennyLane: Automatic differentiation of hybrid quantum-classical computations}''~(2018).
\newblock  \href{http://arxiv.org/abs/1811.04968}{arXiv:1811.04968}.

\bibitem{bp_hea}
N.~Nemkov~(2024).
\newblock  url:~\url{https://github.com/idnm/barren_traps/tree/hea}.

\end{thebibliography}

\end{document}